\documentstyle[11pt,aaspp4]{article}

\input{psfig.sty}

\begin{document}

\title{Timing Noise in SGR~1806-20}

\author{
P.~M.~Woods\altaffilmark{1,2},
C.~Kouveliotou\altaffilmark{1,2},
M.~H.~Finger\altaffilmark{1,2},
E.~{G\"o\u{g}\"u\c{s}}\altaffilmark{2,3},
D.~M.~Scott\altaffilmark{1,2},
S.~Dieters\altaffilmark{2,3},
C.~Thompson\altaffilmark{4},
R.~C.~Duncan\altaffilmark{5}, 
K.~Hurley\altaffilmark{6},
T.~Strohmayer\altaffilmark{7},
J.~Swank\altaffilmark{7}, and
T.~Murakami\altaffilmark{8}
}

\altaffiltext{1}{Universities Space Research Association; 
peter.woods@msfc.nasa.gov}
\altaffiltext{2}{NASA Marshall Space Flight Center, SD50, Huntsville, AL
35812}
\altaffiltext{3}{Department of Physics, University of Alabama in Huntsville, 
Huntsville, AL 35899}
\altaffiltext{4}{Department of Physics and Astronomy, University of North
Carolina, Philips Hall, Chapel Hill, NC 27599-3255}
\altaffiltext{5}{Department of Astronomy, University of Texas, RLM 15.308,
Austin, TX 78712-1083}
\altaffiltext{6}{University of California at Berkeley, Space Sciences
Laboratory, Berkeley, CA 94720-7450}
\altaffiltext{7}{NASA Goddard Space Flight Center, Greenbelt, MD 20771}
\altaffiltext{8}{Institute of Space and Astronautical Science, 3-1-1,
Yoshinodai, Sagamihara-shi, Kanagawa 229, Japan}

\begin{abstract}

We have phase connected a sequence of {\it Rossi X-ray Timing Explorer}
Proportional Counter Array observations of SGR~1806$-$20 covering 178 days.  We
find a simple secular spin-down model does not adequately fit the data.  The
period derivative varies gradually during the observations between 8.1 and 11.7
$\times$ 10$^{-11}$ s s$^{-1}$ (at its highest, $\sim$40\% larger than the long
term trend), while the average burst rate as seen with the Burst and Transient
Source Experiment drops throughout the time interval.  The phase residuals give
no compelling evidence for periodicity, but more closely resemble timing noise
as seen in radio pulsars.  The magnitude of the timing noise, however, is large
relative to the noise level typically found in radio pulsars ($\Delta_8$ = 4.8;
frequency derivative average power $\approx$ 7 $\times$ 10$^{-20}$ cyc$^{2}$
s$^{-3}$).  Combining these results with the noise levels measured for some
AXPs, we find all magnetar candidates have $\Delta_8$ values larger than those
expected from a simple extrapolation of the correlation found in radio
pulsars.  We find that the timing noise in SGR~1806$-$20 is greater than or
equal to the levels found in some accreting systems (e.g., Vela X$-$1,
4U~1538$-$52 and 4U~1626$-$67), but the spin-down of SGR~1806$-$20 has thus far
maintained coherence over 6 years.  Alternatively, an orbital model with a
period $P_{\rm orb}$ = 733 days provides a statistically acceptable fit to the
data.  If the phase residuals are created by Doppler shifts from a
gravitationally bound companion, then the allowed parameter space for the mass
function (small) and orbital separation (large) rule out the possibility of
accretion from the companion sufficient to power the persistent emission from
the SGR.

\end{abstract}


\newpage

\section{Introduction}

The soft gamma repeater (SGR), SGR~1806$-$20 is one of four known SGRs or
sources of brief ($\sim$0.1 s), intense ($\lesssim$ 10$^{42}$ ergs s$^{-1}$)
hard X-ray/soft $\gamma$-ray burst emission (for a review, see Hurley 2000 and
Woods 2000).  All of the SGRs have persistent X-ray counterparts that are
positionally coincident with young ($\sim$ 10$^4$ yr) supernova remnants. Two
SGRs (1806$-$20 and 1900$+$14) are also X-ray pulsars that have been found to 
spin down at a rapid rate (Kouveliotou et al.\ 1998; Hurley et al.\ 1999;
Kouveliotou et al.\ 1999).  The physical interpretation for this rapid
spin-down has been proposed to be magnetic braking of a strongly magnetized
neutron star, or `magnetar' (Kouveliotou et al.\ 1998, 1999).  The magnetar
theory was first developed to explain the extraordinary burst emission from the
SGRs (Duncan \& Thompson 1992; Paczy\'nski 1992; Thompson \& Duncan 1995), and
later extended to include a second class of rare X-ray sources, the so-called
anomalous X-ray pulsars (AXPs; Thompson \& Duncan 1996).

Thompson \& Duncan noted that the AXPs have a number of characteristics (with
the exception of burst emission) that are similar to the SGRs (see Mereghetti
2000 for a review of AXPs).  The AXPs and SGRs have fairly steady X-ray
luminosities ($\sim$ 10$^{35}$ ergs s$^{-1}$).  The persistent spectra of most
AXPs and  SGR~1900$+$14 (Woods et al.\ 1999a) are well represented by a
two-component (blackbody $+$ power-law) model.  The pulse periods of both
groups fall within a narrow range (5 -- 12 s) and are observed to spin down at
rapid rates ($\sim$ 10$^{-12} - 10^{-10}$ s s$^{-1}$).  For three AXPs
(4U~0142$+$61, 1E~2259$+$586, 1E~1048.1$-$5937), the spin-down is dominated by
a secular component, but shows small, yet significant deviations from this
linear trend.  Heyl \& Hernquist (1999) have estimated the noise level within
the frequency histories covering more than 10 years for each of these AXPs. 
They find that the timing noise levels of these AXPs are consistent with an
extrapolation of the correlation found between timing noise and spin-down rate
present in radio pulsars (Arzoumanian et al.\ 1994).  Kaspi, Chakrabarty \&
Steinberger (1999) have recently analyzed an extended sequence of {\it Rossi
X-ray Timing Explorer (RXTE)} Proportional Counter Array (PCA) observations of
two AXPs (1E~2259$+$586 and 1RXS~J1709$-$40).  For each source, they were able
to phase-connect the data over a two year time span.  They found that the
spin-down was fairly constant, although, the residuals did show some marginally
significant timing noise (i.e.\ a third-order phase term).

Similar to the aforementioned AXPs, recent work has shown that SGR~1900$+$14
does not spin down at a constant rate either (Kouveliotou et al.\ 1999; Woods
et al.\ 1999a; Marsden et al.\ 1999; Woods et al.\ 1999b).  In the magnetar
model, the SGRs are seismically active neutron stars, and there are a few
possible sources of spindown variations.  The relative strengths of the
conduction current and the displacement current in the outer magnetosphere will
be  modified by bursting activity:  both by direct ejection of particles, and
by the rearrangement of the surface magnetic field (Thompson et al.\ 1999). 
The resulting increase in the spindown torque could be significant for a slowly
rotating neutron star,  if the magnetic dipole were (approximately) aligned
with the rotation axis.  Alternative possibilities include enhanced angular
momentum loss due to persistent emission of  Alfv\'en waves and particles
(Thompson \& Blaes 1998;  Harding,  Contopoulos, \& Kazanas 1999); and
long-period precession driven by the asymmetric inertia of the corotating
magnetic field (Melatos 1999) or by crustal fractures (Thompson et al.\ 1999).

Alternatively, the variable spindown of the SGRs has been ascribed to an
enhanced propeller or accretion torque acting on more conventional magnetic
fields ($B_{\rm dipole} \sim 10^{11}-10^{12}$ G), by several authors.   Marsden
et al.\ (1999, 2000) suggest that the SGRs are neutron stars born with large
kick velocities in a dense inter-stellar medium (ISM).  In such a situation,
the neutron star may catch up with the slowing ejecta, but accretion at the
rate inferred from the luminosities of these sources requires a small relative
motion ($< 10$ km s$^{-1}$).  Van Paradijs et al.\ (1995) earlier proposed that
the AXPs are surrounded by fossil disks, left over from the evolution of 
Thorne-\.Zytkow objects (T\.ZO).  The T\.ZOs are compact objects that have
entered the stellar envelope of their companion and spiraled into the center
(Thorne \& \.Zytkow 1977).  A variant of this model, involving fallback disks
formed during the early dynamical evolution of supernova, was recently applied
to the SGRs by  Chatterjee, Hernquist, \& Narayan (1999) and Alpar (1999). 
While these scenarios plausibly account for the observed timing noise, the
relation between accretion and the hyper-Eddington SGR flares (in particular
the intense $L\sim 10^7\,L_{\rm Edd}$ gamma-ray spikes of the giant flares)
remains largely mysterious.

Until now, the spin history of SGR~1806$-$20 was composed of three widely
spaced period measurements covering three years (Kouveliotou et al.\ 1998). 
Here, we have phase connected a long sequence of {\it RXTE}-PCA observations
between 1999 February 12 and 1999 August 8.  We find that superposed on the
dominant quadratic trend in the phases (spin-down term) are significant
residuals, i.e., timing noise.  We quanitify the level of timing noise and
compare it to the levels found in AXPs, magnetically braking radio pulsars, and
accreting pulsars.  We place limits on any periodicities in the phase residuals
and constrain the orbital parameters of any potential companion.

\section{Timing Analysis}

\begin{figure}[!htb]
\centerline{
\psfig{file=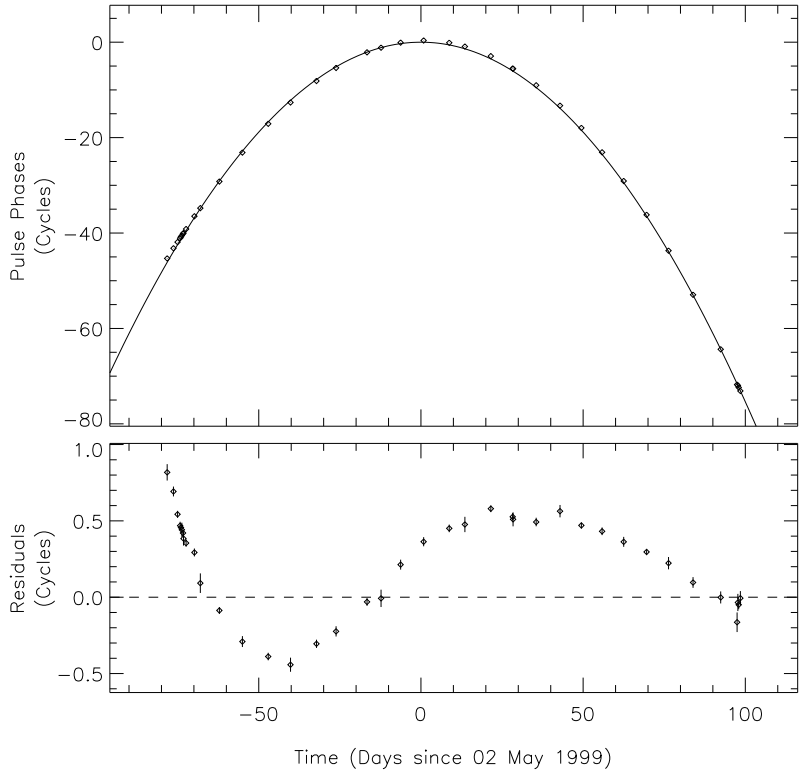,height=4.5in}}
\vspace{-0.15in}
\caption{Pulse phases for SGR~1806$-$20 during 1999 minus a linear trend 
{\it (top)} and minus a quadratic {\it (bottom)}.}
\vspace{11pt}
\end{figure}

Between 1999 February 12 and 1999 August 8, SGR~1806$-$20 was routinely
observed with the PCA aboard {\it RXTE} for a total of 322 ks.  The monitoring
campaign started with a 50 ks exposure, followed by twenty-two 10 ks
observations whose spacing grew from 0.5 days to 7 days where it remained until
August.  Near the end of the observing sequence, the intervals between
observations gradually became shorter and terminated with another 50 ks
observation.  For each observation, we avoided intervals with clear burst
emission, extracted 2 -- 10 keV photons, and binned them at 0.125 s time
resolution.  We barycenter corrected the bin times and performed an epoch-fold
search (7.480 s $-$ 7.485 s) on the data about the period predicted from the
spin-down measured by Kouveliotou et al.\ (1998).  The initial $\sim$50 ks of
data were folded on the frequency with the largest $\chi^2$ value from the
epoch-fold search in order to create a template profile.  We then folded
individual 3 -- 11 ks segments of the same data on this frequency.  A fast
Fourier transform was applied to the folded profiles of both the individual
segments and the net template profile.  Using the first four harmonics of the
Fourier representation of the profiles, the individual profiles were
cross-correlated with the template profile yielding the relative phase, and
intensity of each segment.  A new ephemeris was obtained by fitting these
phases to a polynomial of the form $\phi(t) = \phi_{0} + \nu \left( t - t_{0}
\right) + \frac{1}{2} \dot{\nu} \left( t - t_{0} \right) ^2 +
\frac{1}{6}\ddot{\nu} \left( t - t_{0} \right) ^3 + \dots + \frac{1}{n!}
\nu^{(n)} \left( t - t_{0} \right) ^n$, where $\phi(t)$ is the phase, $\phi_0$
is a phase offset, $t_0$ is the epoch, $\nu$ is the frequency, $\dot{\nu}$ is
the frequency derivative, $\ddot{\nu}$ is the second time derivative of the
frequency, and $\nu^{(n)}$ is the $n^{\rm th}$ derivative of the frequency. 
This procedure was iterated, each time using a revised template profile
determined from the previous data set.  As more data were gradually included in
the analysis, the addition of higher order polynomials became necessary to
obtain an adequate fit to the phases (see Figure 1).  When the entire data set
was fit, we were able to determine the frequency and the first three
derivatives (the detection of the 4$^{\rm th}$ frequency derivative is
marginal; see Table 1 for all fitted parameters).  Throughout this sequence of
observations, the pulsed intensity remained constant.

\placetable{tbl-1}

In order to directly compare with the 1996 PCA observations of SGR~1806$-$20
(Kouveliotou et al.\ 1998), we applied the same technique to the 1996 data and
measured a frequency and frequency derivative (see Table 1).  These values are
consistent with those found by Kouveliotou et al.\ (1998).  The temporal
baseline of these data is much shorter (13 days) than the 1999 observations,
and so, timing noise of the same magnitude as that found in 1999 is
undetectable.  The pulse profile (2 -- 10 keV) generated from the 1996
observations is very similar to the 2 -- 24 keV profile shown in Kouveliotou et
al.\ (1998), having a single broad peak with a narrow valley.  The pulse
profile in 1999 is similar in that it only shows one peak, however, the width
of the valley at this epoch is slightly broader than the peak.

\begin{figure}[!p]
\centerline{
\psfig{file=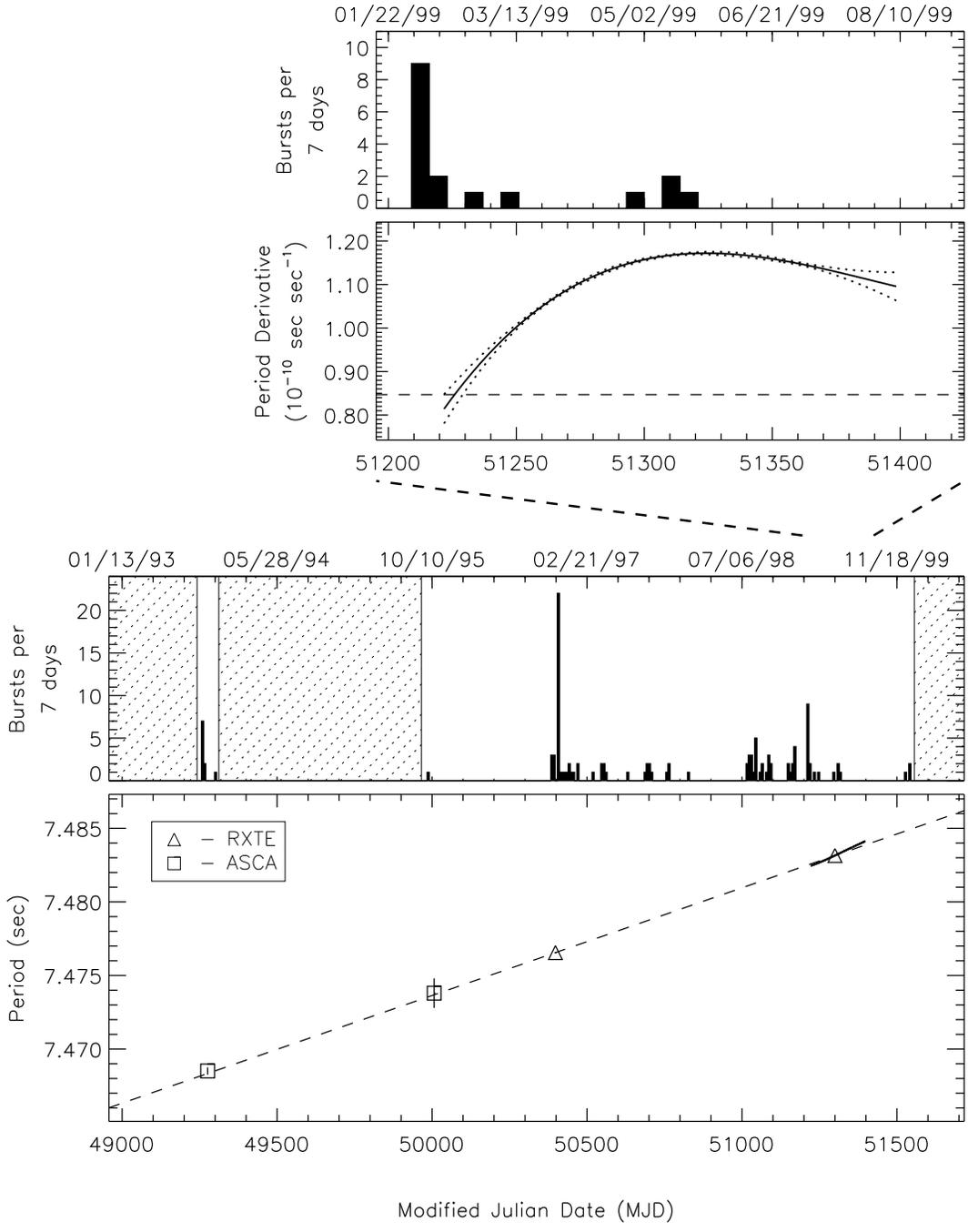,height=7.5in,%
bbllx=54bp,bblly=72bp,bburx=594bp,bbury=726bp,clip=}}
\vspace{-0.15in}

\caption{{\it Bottom} -- Burst rate history {\it (upper panel)} and period
history {\it (lower panel)} of SGR~1806$-$20 from 1993 September through 1999
August.  Lower axis label is Modified Julian Date and upper axis is mm/dd/yy. 
Shaded regions denote intervals where an off-line burst search was not
performed and no triggered events were recorded.  The dashed line indicates a
least squares fit to the period measurements.  {\it Top} -- Inset of lower
figure showing burst rate history {\it (upper panel)} and period derivative
history {\it (lower panel)} of SGR~1806$-$20 during the 1999 observations with
the PCA.  The dotted lines represent $\pm 2\sigma$ errors on the period
derivative.  The dashed line is the long-term spin-down rate.}

\end{figure}

Combining our results with the {\it ASCA} results reported in Kouveliotou et
al.\ (1998), we construct a period history of SGR~1806$-$20 over 6 years
(Figure 2).  We have also performed an off-line search for untriggered events
in the BATSE data using the methodology described in Woods et al.\ (1999b). 
Plotted along with the period history is the burst rate history as seen with
BATSE (Figure 2).  To demonstrate the non-uniformity of the period derivative,
we plot an exploded view of the period derivative versus time during the 1999
observations.

\begin{figure}[!htb]
\centerline{
\psfig{file=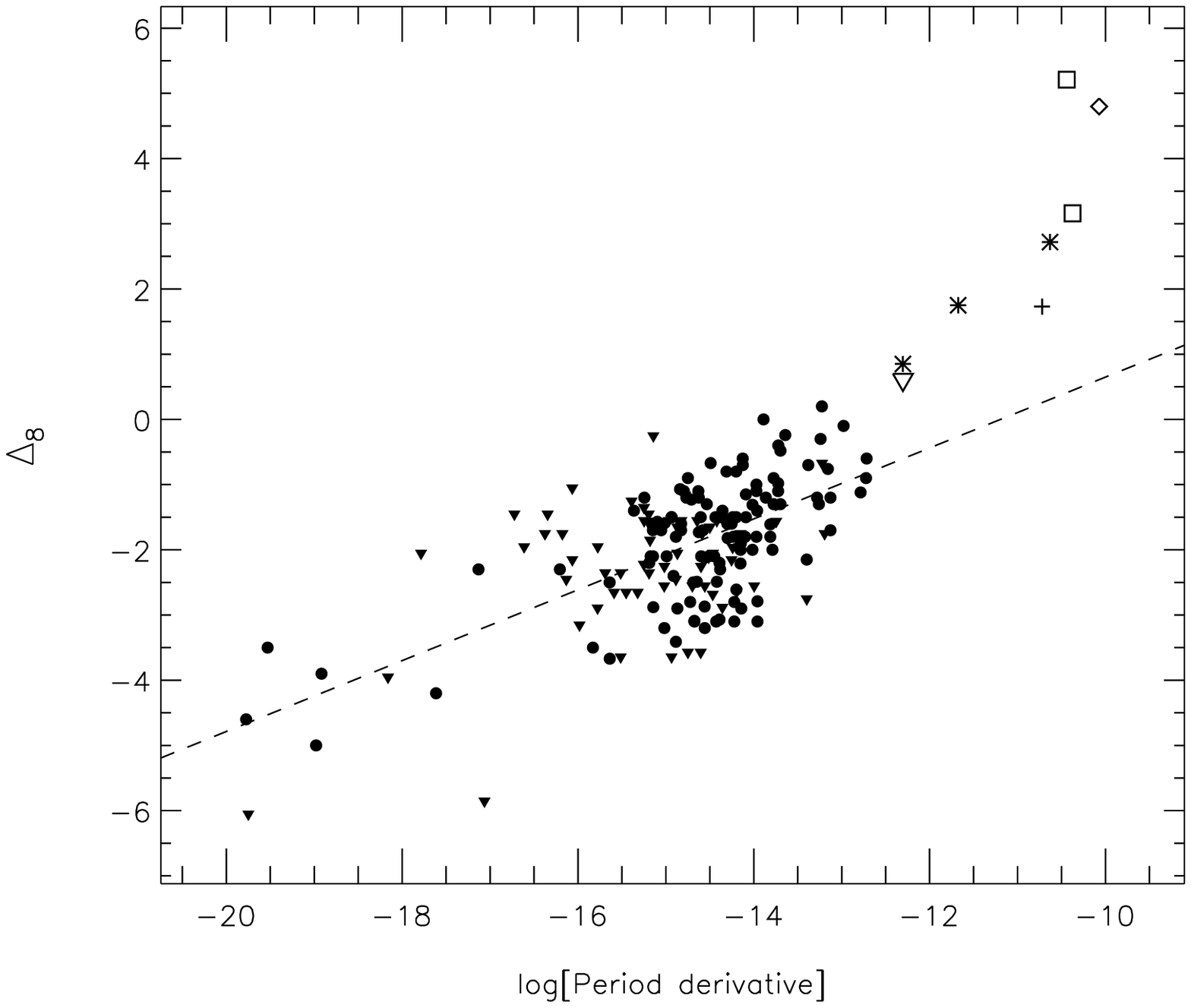,height=4.5in}}
\vspace{-0.05in}

\caption{The timing noise parameter, $\Delta_8$, plotted versus the period
derivative for 139 radio pulsars, 2 accreting X-ray pulsars, 4 AXPs, and one
SGR.  The filled circles are measurements of $\Delta_8$ and the inverted,
filled triangles are upper limits, both for the radio pulsars (Arzoumanian et
al.\ 1994).  The asterisks denote measurements of $\Delta_8$ from frequency
measurements of three AXPs (Heyl \& Hernquist 1999). The inverted, open
triangle denotes a 2$\sigma$ upper limit of $\Delta_8$ for 1E~2259$+$586, and
the plus sign is the $\Delta_8$ value of 1RXS~J1709$-$40 based upon a 4$\sigma$
detection of $\ddot{\nu}$ (Kaspi et al.\ 1999).  The diamond marks $\Delta_8$
for SGR~1806-20.  For comparison, $\Delta_8$ values of two accreting systems,
GX~1$+$4 and 4U~1626$-$67 (Chakrabarty 1996), are also shown (squares).  The
dashed line represents a least squares fit to the radio pulsar measurements
only (i.e.\ excluding upper limits).  Please note that 43 radio pulsars (see
Arzoumanian et al.\ 1994) and one AXP (1E~2259$+$586) have duplicate entries.}

\vspace{11pt}
\end{figure}

Motivated by the techniques applied to other pulsars in estimating timing
noise, we calculated a power-density spectrum of the frequency derivative
residuals and the $\Delta_8$ value for SGR~1806$-$20.  Using the method
described by Deeter \& Boynton (1982) for unevenly sampled data, we obtained a
spectrum for the frequency derivative residuals.  We detect significant power
in the range 0.3 -- 4 $\times$ 10$^{-7}$ Hz at an average level of $\approx$ 7
$\times$ 10$^{-20}$ cyc$^{2}$ s$^{-3}$ and consistent with being a flat
spectrum.  An alternative method to estimate the level of timing noise in a
pulsar is to calculate the $\Delta(t)$ parameter ($\Delta(t) \equiv
\log{\frac{\vert\ddot{\nu}\vert t^3}{6\nu}}$ [Arzoumanian et al.\ 1994]).  In
order to compare our results to those of Arzoumanian et al., we calculated
$\Delta_8$ or $\Delta(10^8)$ for SGR~1806$-$20.  Taking the average
$\ddot{\nu}$ over the span of the data ($\langle\vert\ddot{\nu}\vert\rangle
\approx$ 5.1 $\times$ 10$^{-20}$ Hz s$^{-2}$), we find $\Delta_8$ = 4.8 for
SGR~1806$-$20.  Figure 3 displays the $\Delta_8$ values for 139 radio pulsars
(Arzoumanian et al.\ 1994), 4 AXPs (Heyl \& Hernquist 1999; Kaspi et al.\
1999), 2 accreting X-ray pulsars (Chakrabarty 1996), and SGR~1806$-$20.  All
measurements of $\Delta_8$ for the AXPs and SGR~1806$-$20 are positioned above
an extrapolation of the trend found for the radio pulsars.  The 2$\sigma$ upper
limit of 1E~2259$+$586 from Kaspi et al.\ is consistent with the trend.

If the timing noise is instead a product of Doppler shifts due to a binary
companion, we can fit the phase residuals to an orbit.  To do so, we assume a
constant period derivative equal to the long-term trend given by a fit to the
period measurements ($\dot{P} \sim$ 8.47 $\times$ 10$^{-11}$ s s$^{-1}$).  For
this analysis, we used both the phases from the 1996 and 1999 {\it RXTE}
observations as well as the frequency measurements from the earlier {\it ASCA}
observations given by Kouveliotou et al.\ (1998).  We fit these data to an
array of orbits with fixed periods between 10 and 5000 days, allowing for a
cycle slip between the separate observations.  The reduced $\chi^2$ reaches a
minimum value of 1.3 at an orbital period $P_{\rm orb}$ = 733 $\pm$ 14 days. 
For this best fit period, we find a mass function $f(M)$ $\approx$ 8.8 $\times$
10$^{-2}$ $M_{\odot}$ and an orbital separation $A_{\rm x} \sin{i}$ $\approx$
360 lt-s.  There is also a secondary minimum in $\chi^2$ at $P_{\rm orb}
\approx$ 380 days ($\chi^2_{\nu}$ = 1.7) that is only marginally less favorable
than the 733 day orbit.

\section{Discussion}

We have shown that SGR~1806$-$20 exhibits strong timing noise during its rapid
spindown.  The amplitude of the torque variations is significantly higher than
in recent measurements of the AXPs 1E~2259$+$586 and  1RXS~J1709$-$40 by Kaspi
et al.\ (1999), and lies well above the trend established by Arzoumanian et
al.\ (1994) for radio pulsars.  The noise could, in principle, be caused by
orbital Doppler shifts, variations in magnetospheric current flows driven by
magnetic activity in a magnetar, stellar precession,  or variations in the
torque from an accretion disk.  The noise in  SGR~1806$-$20 does not appear to
be predominantly due to glitches. 

If the spin frequency variations of SGR~1806$-$20 are due to a gravitationally
bound companion, then the stars must be widely separated and the companion mass
($M_{\rm c}$) must be small (0.74 $M_{\odot} < M_{\rm c} <$ 1.7 $M_{\odot}$ for
an inclination angle $i >$ 10$^{\circ}$ and an assumed 1.4 $M_{\odot}$ neutron
star).  For this particular orbital solution, accretion from the companion can
be excluded as the energy source powering the persistent X-ray emission (i.e.,
$\dot{M} \ll 10^{15}$ g s$^{-1}$).  Alternatively, one may conjecture that the
SGR is accreting from a circumstellar disk (Chatterjee et al.\ 1999; Alpar
1999), or perhaps from co-moving ejecta/ISM material (Marsden et al.\ 2000). 
The magnitude of the timing noise in SGR~1806$-$20 is larger than any known
radio pulsar, yet falls within the boundaries of accreting X-ray pulsars (see
Figure 2).  Over a similar frequency range (see Table 5 of Bildsten et al.\
1997), the accreting systems Vela~X$-$1, 4U~1538$-$52 and 4U~1626$-$67 have
average power levels of 2.2, 1.7, and $\sim$0.1 $\times$ 10$^{-20}$ cyc$^2$
s$^{-3}$.  Accretion models can account for the frequency derivative variations
as due to torque fluctuations from interactions between the stellar magnetic
field and the surrounding material.  Torque fluctuations may then lead to
variations in the source luminosity (Ghosh \& Lamb 1979; see also Bildsten et
al.\ 1997 for observational evidence that suggests the torque/luminosity
relationship may be more complicated), however, the pulsed intensity of
SGR~1806$-$20 remained constant throughout this interval.  Furthermore, an
important difference between these accreting systems and SGR~1806$-$20 is that
the accreting systems have shown extended intervals of spin-up (Bildsten et
al.\ 1997), whereas SGR~1806$-$20 and all other SGRs and AXPs have not.

Variations in spin-down are not, however, limited to accreting neutron stars.
Timing noise is present in isolated radio pulsars, and is strongest in the
youngest members of that population (Cordes \& Helfand 1980; Arzoumanian et
al.\ 1994).  One of the key premises of the magnetar model is that the SGRs are
young and seismically active, with their recurrent outbursts being triggered by
energetic fractures of the neutron star crust (Thompson \& Duncan 1995). 
Seismic activity in any magnetized neutron star can modify the external torque
through the production of an outward flowing relativistic wind (Thompson \&
Blaes 1998); or by increasing the conduction current relative to the
displacement current in the outer magnetosphere (Thompson et al.\ 1999).  For
example, a mass as large as $\Delta M \sim B_{dipole}^2 R_{NS}^6 
\Omega^{4/3}/4\pi(GM_{NS})^{5/3} = 2\times 10^{20}\, (B_{dipole}/4\times
10^{14}~{\rm G})^2\,(P/8~{\rm s})^{-4/3}$ g can be suspended in the outer
magnetosphere by centrifugal forces, and can easily be supplied through
hyper-Eddington bursting activity.  It is not surprising then to find a high
level of timing noise in SGR~1806$-$20, compared with radio pulsars. 
Furthermore, the strength of the timing noise in SGR~1806$-$20  relative to the
AXPs provides a strong hint that torque variations in an isolated,
highly-magnetized neutron star are correlated, perhaps indirectly, with
activity as a burst source.

Long period precession has been suggested as a source  of spindown variations
in magnetars.  Melatos (1999) proposed that precession would be driven by the
asymmetric inertia of the external magnetic field, coupled to the 
hydromagnetic distortion of the star.  Alternatively, free precession (of a
lower amplitude) could be excited by bursting activity (Thompson et al.\ 1999).
However, a precession period $\tau_{\rm pr}$ requires that the moment of
inertia of the {\it pinned} crustal superfluid does not exceed 
$I_{pinned}/I_{NS} \sim  P/\tau_{\rm pr} =  3\times 10^{-7}\,(\tau_{\rm
pr}/1~{\rm yr})^{-1}\,(P/8~{\rm s})$ (Shaham 1977), several orders of magnitude
smaller than is inferred for young, glitching pulsars.  Thus far, there is no
compelling evidence that shows the observed spin-down variations are periodic. 
Further monitoring of the pulse frequency is required to demonstrate otherwise.

\acknowledgments{\noindent {\it Acknowledgements} -- We thank the RXTE SDC for
pre-processing the RXTE data.  We also thank J. Heyl and Z. Arzoumanian for
providing the AXP and radio data, respectively.  We thank the referee J. Heyl
for useful comments.  PMW, CK, MHF, EG, and DMS acknowledge support from the
cooperative agreement NCC 8-65.  CT acknowledges support from the Alfred P.
Sloan Foundation and NASA grant NAG 5-3100.  RD acknowledges support from both
the Texas Advanced Research Project grant ARP-028 and NASA grant NAG 5-8381. }

\begin{center}
\begin{deluxetable}{lcc}
\small
\tablecaption{Pulse ephemeris for SGR~1806$-$20 from {\it RXTE} PCA 
observations. \label{tbl-1}}
\tablewidth{6.5in}

\tablehead{
\colhead{Parameter}     &
\colhead{1996 November 05 -- 18}     &
\colhead{1999 February 12 -- August 8}     \\
}

\startdata

Epoch (MJD)                            &  50398  &  51300  \nl 

Exposure (ksec)                        &  133.5  &  322.2  \nl 

$\chi^2$/dof                           &  1.8/8  &  55.4/33  \nl 

$\nu$ (Hz)                             &  0.133751469(20) & 0.133633498(5)  \nl 

$\dot{\nu}$ (10$^{-12}$ Hz s$^{-1}$)   &  -1.22(17)  &  -2.0666(27)  \nl 

$\ddot{\nu}$ (10$^{-20}$ Hz s$^{-2}$)  &  ...  &  -2.64(19)  \nl 

$\nu^{(3)}$ (10$^{-26}$ Hz s$^{-3}$)   &  ...  &  1.47(7)  \nl 

$\nu^{(4)}$ (10$^{-33}$ Hz s$^{-4}$)   &  ...  &  -1.9(6)  \nl 

$P$ (s)                                &  7.4765534(12)  &  7.48315368(29)  \nl 

$\dot{P}$ (10$^{-11}$ s s$^{-1}$)      &  6.8(10)  &  11.572(15)  \nl 

\enddata

\end{deluxetable}
\end{center}

\end{document}